# Support for Evolving Software Architectures in the ArchWare ADL

[1]Ron Morrison, [1]Graham Kirby, [1]Dharini Balasubramaniam, [1]Kath Mickan, [2]Flavio Oquendo,
[2]Sorana Cîmpan, [3]Brian Warboys, [3]Bob Snowdon, [3]R Mark Greenwood
[1]*School of Computer Science, University of St Andrews, St Andrews, Fife KY16 9SS, UK*
[2]*ESIA, Université de Savoie, 5 Chemin de Bellevue, 74940 – Annecy-le-Vieux, France*
[3]*Department of Computer Science, University of Manchester, Manchester M13 9PL, UK*
*{ron, graham, dharini, kath}@dcs.st-and.ac.uk*
*{Flavio.Oquendo, Sorana.Cimpan}@esia.univ-savoie.fr*
*{brian, rsnowdon, markg}@cs.man.ac.uk*

**Abstract**

*Software that cannot evolve is condemned to atrophy: it cannot accommodate the constant revision and re-negotiation of its business goals nor intercept the potential of new technology. To accommodate change in software systems we have defined an active software architecture to be: dynamic in that the structure and cardinality of the components and interactions are changeable during execution; updatable in that components can be replaced; decomposable in that an executing system may be (partially) stopped and split up into its components and interactions; and reflective in that the specification of components and interactions may be evolved during execution.*

*Here we describe the facilities of the ArchWare architecture description language (ADL) for specifying active architectures. The contribution of the work is the unique combination of concepts including: a π-calculus based communication and expression language for specifying executable architectures; hyper-code as an underlying representation of system execution that can be used for introspection; a decomposition operator to incrementally break up executing systems; and structural reflection for creating new components and binding them into running systems.*

## 1 Introduction

Software architectures [1, 2] describe systems in terms of their components and interactions between components. We define an active software architecture to be: dynamic in that the structure and cardinality of the components and interactions are changeable during execution; updatable in that components can be replaced dynamically; decomposable in that an executing system may be (partially) stopped and split up into its components and interactions; and reflective in that the specification of components and interactions may be evolved during execution.

Active architectures address problems of co-evolution in dynamically changing commercial environments where business changes create pressures on the software to evolve, and at the same time technology changes create pressures on the business to evolve. The business effects of introducing, or changing, such software systems are often emergent and require their software architecture models to accommodate their demands by being dynamically evolvable themselves.

Often evolution is achieved by taking the system offline, editing and re-compiling the source code, and finally rebinding or starting up the new system. In many cases, especially in long-lived systems, such an approach is infeasible since: the source code may no longer be available; there may be valuable data encapsulated in the closure of executing components; the cost of assembly may be prohibitive; or down-time may not be an option. Examples of such systems include: continuously running business process models; autonomic systems; GRID applications; self-adapting/tuning systems; peer-to-peer routing systems; control systems; and pervasive computing applications.

Figure 1 shows an evolving system. At the initial stage (a), the system is composed of three components of one kind (say clients) interacting with one component of another kind (say server) that has access to some data. At stage (b), this system has been decomposed to yield the individual components with the server still maintaining its access to the data. The next stage (c) sees the components evolved so that we have three clients and two servers both of which maintain the access to the shared data. Finally at stage (d) a new evolved system is formed by composing the five components so that one client interacts with one server and the other two clients interact with the other server.

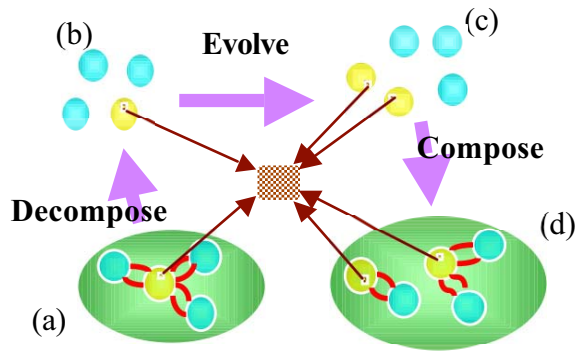

**Figure 1. An evolving system**

The essence of an active architecture is that it encompasses the evolution of the architectural specification (model) as well as the evolution of the enacted system. Our claim is that active architectures require a paradigm shift in ADLs.

This paper describes the ArchWare Architecture Description Language [3], which is sufficiently rich to provide executable specifications of active systems. We define a core language on which architectural styles can be layered and on which a construction methodology can be applied. Our focus is on the base technologies required to support dynamic and evolvable systems including: a π-calculus based communication and expression language for specifying executable architectures; hyper-code as an underlying representation of system execution that can be used for introspection; a decomposition operator to incrementally break up executing systems; and structural reflection for creating new components and binding them into running systems. The contribution of the work is the unique combination of these concepts.

## 2 Related work

Hitherto researchers have proposed many formal ADLs for representing and analysing architectural designs. Adage [4] supports the use of architectural frameworks in the avionics industry; Aesop [5] has architectural styles; MetaH [6] has specific guidance for real-time avionics control software; SADL [7] provides a formal basis for architectural refinement; C2 [8] supports the description of user interface systems; Wright [9] supports the specification and analysis of interactions and UniCon [10] supports a mixture of heterogeneous component and connector types.

Gerel [11] accommodates changes that are robust to system evolution. It supports the definition of generic programmed changes and provides for changes that are only applied when a precondition is satisfied by the current configuration.

LEDA [12] is an ADL based on the π-calculus which aims to address problems of refinement and validation by combining formal methods with object-oriented concepts. Systems can be checked for compatibility before being composed. Re-use is encouraged by polymorphism of behaviours, which allows architectures to be parameterised.

Containment Units [13] provide a mechanism for dealing with anticipated change. ArchStudio [14] is a tool suite that supports architecture-based development. Changes are made to an architectural model and then reified into implementation by a runtime architecture infrastructure. In [15], specific component managers identify external architecture changes by listening to events, and then react in order to preserve architecture constraints. The constraints themselves cannot be evolved.

Darwin [16] is a declarative binding language for building distributed systems, which is based on the π-calculus. It has a similar structure to the ArchWare ADL but a more limited approach to evolution. In Darwin, evolution is restricted to the creation of new bindings between components and the instantiation of new components. Dynamic instantiation is either lazy or direct. In lazy instantiation, instances of a component are created, as they are needed in the computation. In direct instantiation, a component's context is statically defined, but the component itself can be defined in a 'meta-level configuration', essentially a Darwin script, and instantiation driven by interpreting this script at run-time.

ArchJava [17] is an extension to Java that unifies a system's software architecture and implementation. The type system ensures that implementation code conforms to architectural constraints. It provides for dynamic architectures only in that statically defined components can be dynamically instantiated and connected.

Each ADL has its own focus according to needs and taste with little integration of the overlapping ADL concepts. ACME [18] is an attempt at such integration but does it at the level of a lowest common denominator.

Most work on ADLs concentrates on the specification and enactment model where the formal properties of the system are specified, analysed and then executed. Where dynamic evolution is considered, its scope is limited and it is treated as distinct from the initial system development. The focus of the ArchWare project [19] is in evolving systems, with emergent properties, where the system is executing continuously. The model specification and the model enactment are both regarded as part of a single executing state. At any point in time the model specification will be an accurate description of the model execution.

Our work on hyper-code intersects Intentional Programming [20], which creates a programming environment based on a graph representation of the program. Compilation invokes generators that operate over the graph to assemble the program.

## 3 Change in active architectures

We have identified three architectural kinds of change in active architectures. These are:

- Dynamic change: allows the topology of the components and interactions to be changed dynamically. New components and interactions may be created during execution.
- Update change: allows components to be replaced.
- Evolutionary change: allows the specification of the components and interactions to be changed during execution.

There are three main stages involved in evolving active architectures:
- deciding when changes are required; all changes are made in response to some stimulus
- deciding what changes are required in reaction to the stimulus and the environment
- applying the change via an appropriate change mechanism

Taken from the field of control systems, Figure 2 shows how application knowledge, obtained by measurement perhaps, is used to achieve a goal, causing a reaction. The reaction may be to continue execution or to initiate some change mechanism.

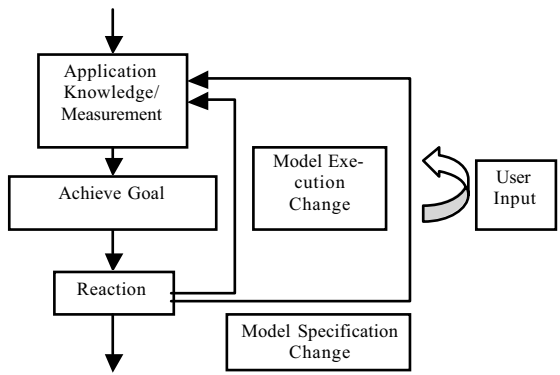

**Figure 2. Change mechanisms**

In an active architecture the specification of the architectural model changes with the model execution. Thus changes during execution will change the specification and changes to the specification will affect the execution. However at any time in the execution of the model the specification is dynamically up-to-date.

In the ArchWare ADL we have provided a number of change mechanisms reflecting our estimate of the frequency of the expected type of change. Dynamic change and update change are made using mechanisms built into the specification language. For example, this may be adding more components to execute in parallel or passing components to other components through connections (dynamic change) or replacing a component from a library of parts by assignment (update change).

Evolutionary change is characterised by changing specifications and requires a reflective system. That part of the system to be changed is stopped, its specification altered and the new specification enacted in the executing model. Evolutionary change may be implemented using introspection on the component to be changed to yield its specification, and reflection to rebind the new specification. The reflection itself can use existing specifications to make the alterations. In the limit user input may be necessary to accommodate unpredicted emergent behaviour.

## 4 ArchWare overview

This work is undertaken within the EC funded ArchWare software architecture framework. The ArchWare project takes a holistic view of software development. Its aims are to advance and integrate research on software architecture and reflective systems to develop languages, frameworks and tools for architecting and engineering dynamic and evolvable software systems. The important aspects of the ArchWare approach can be described as follows:
- a formal, style-based, executable architecture description language to describe architectural structure, behaviour, qualities and evolution of systems
- a suite of tools based on the ADL for architecture design and analysis
- run-time and environment framework to support the development and deployment of software systems and coordination of design and analysis tools
- generic and customisable process models for evolutionary, architecture-centric development of software systems

## 5 The ArchWare Style Language

The ArchWare style language provides a framework for formalising software architectures based on the concepts of components and connectors. It is used to define families of architectures that have common structure and satisfy the same properties.

The style language makes use of both the underlying ArchWare ADL (see Section 6) and the ArchWare AAL (Architecture Analysis Language). The structure and behaviour of an architecture family are specified using the ADL while the constraints on the family are specified using the AAL.

Styles in ArchWare are defined as property-guarded abstractions that may be applied to yield instances of an architecture conforming to the style.

A style definition consists of the following parts:
- types
- constituent elements
- constraints
- analysis

Figure 3 shows a specification of a Client-Server style. *Client_Server* defines three elements to be used in its specification. *Client* and *Server* are component styles while *PC* (procedure call) is a connector style. Three constraints are specified for *Client_Server*:
- a components can only be a *Client* or a *Server*
- a *Client* can only be connected to a *Server*

- a *Server* can only be connected to a *Client*

In the analysis part, *connected* checks whether two components are connected.

```
Client_Server is style where {
elements
  Client is style extending Component
  …
  Server is style extending Component
  …
  PC is style extending Connector
  …
constraints
  to connectors apply {
    forall(c|c in style PC)}.
  to components apply {
    forall(c|c in style Client
    or c in style Server),
    forall(c1,c2|c1 connected to c2
      implies
        (c1 in style Client and
        c2 in style Server) or
        (c1 in style Server and
        c2 in style Client)
  };
analysis
  connected is AAL_property
  parameters
    c1 in style Component,
    c2 in style Component;
  property
  to connectors apply {
    exists(conn | c1 attached to conn
      and c2 attached to conn )
  }
}
```

**Figure 3: Client/Server Styles**

The architecture styles can be used to specify evolution but can be also be part of the evolution of another component. The theorem provers and model checkers that ensure correctness must be integrated within the evolutionary process to spark change when a constraint is violated, an issue for current research. Here we concentrate on the required environment and base technologies within the ArchWare ADL for dynamic expression and evolution. These include the following:
- a formal foundation based on higher-order π-calculus [21] for specifying components of architectures with dynamic structure and cardinality
- integration of a π-calculus based language for communication and an expression based language to yield executable specifications
- hyper-code as a representation for system execution to support reification
- a *decompose* operator essential to break up active systems into their components prior to evolution and recomposition
- structural reflection to support evolution

The theoretical foundation enables formal analysis of the architecture and proof of its desired properties. Executable specifications reduce the cost and complexity of separately implementing the corresponding software systems. The separation of concerns of co-ordination, communication and computation of components make the system easier to understand and evolve. Dynamic expression is built into the constructs of the language. The facilities for composition and reification together with support for decomposition and reflection enable evolving systems.

Formal analysis of architectures in the ADL is performed at the style layer. The styles specify the desirable properties of the architecture and theorem provers and model checkers are used to verify these properties. It is important to note that the formal checking may be applied to a static description incrementally to a trace of the executing system. However, we concentrate here on the facilities for evolution in the ADL.

## 6  The ArchWare ADL

The ArchWare ADL is the simplest of a family of languages designed for modelling active software architectures based on the concepts of π-calculus, persistent programming and dynamic system composition and decomposition.

The ArchWare ADL is a strongly, and mostly statically, typed persistent language. The ADL system consists of the language and its populated persistent environment and uses the persistent store to support itself. To model the component and communication algebra, the ADL supports the concepts of behaviours, abstractions of behaviours and connections between behaviours. Communication between components, represented by behaviours, is via channels, represented by connections. For expressing data the language also supports a number of data types: *integer*, *boolean*, *real*, *string*, *locations*, *views*, *sequences* and *higher order functions*. These can be regarded as syntactic sugar since they can all be encoded in the π-calculus. The language also supports all the basic π-calculus and expression based operations as well as composition and decomposition.

The ArchWare ADL is designed using the three principles of abstraction, correspondence and type completeness [22-24].
- The principle of *abstraction* allows abstractions over every semantically meaningful syntactic category in the language. Thus functions are abstractions over expressions.
- The principle of *correspondence* states that the rules for introducing and using names should be the same throughout. In particular there should be a one-to-one correspondence between introducing names in declarations and as parameters.
- The principle of *type completeness* states that the rules for using data types must be complete with no gaps. For example, general rules for type constructors should have no exceptions.

The application of these design rules yields languages that are both small in the number of concepts and power-ful. They are small in that there are no exceptions to the rules and powerful since every combination is valid. These properties are important in the design of hyper-code and its programmable interface (see later in Section 7).

The ArchWare ADL is designed using a layered approach with the above-mentioned principles of programming language design guiding the process. Layering the language helps to separate different concerns. There are currently three layers in the ArchWare ADL:

- The base layer defines a coordination language without data values corresponding to non-higher-order monadic π-calculus.
- The first order layer adds data values and abstractions and corresponds to non-higher-order polyadic π-calculus.
- The higher order layer corresponds to higher-order polyadic π-calculus. We present this layer in this paper.

## 6.1 The ArchWare ADL type system

The ArchWare ADL type system is based on the notion of types as a set structure imposed over the value space. Membership of the type sets is defined in terms of common attributes possessed by values, such as the operations defined over them. These sets or types partition the value space. They may be predefined, like *integer*, or they may be formed by using one of the predefined type constructors such as *view*.

The constructors obey the principle of type completeness. That is, where a type may be used in a constructor, any type is legal without exception. This has two benefits. Firstly, since all the rules are very general and without exceptions, a very rich type system may be described using a small number of defining rules. This reduces the complexity of the defining rules. Secondly the type constructors are as powerful as is possible since there are no restrictions on their domain.

The universe of discourse of the ArchWare ADL can be described as follows. The following base types are defined:

1. The scalar data types are *integer*, *real*, and *boolean*.
2. Type *string* is the type of a character string; this type embraces the empty string and single characters.
3. Type *any* is an infinite union type; values of this type consist of a value of any type together with a representation of that type.
4. Type *behaviour* is the type of an executing process.

The following type constructors are defined:

5. For any type $t$, *location [t]* is the type of a location that contains a value of type $t$.
6. For any type $t$, *sequence[t]* is the type of a sequence with elements of type $t$.
7. For identifiers $I_1,...,I_n$ and types $t_1,...,t_n$, *view[$I_1$: $t_1$,...$I_n$: $t_n$]* is the type of a view with fields $I_i$ and corresponding types $t_i$, for $i = 1..n$ and $n \geq 0$.
8. For any types $t$ and $t_1,...,t_n$, *function[$t_1,...,t_n$] $\rightarrow$ t* is the type of a function with parameter types $t_i$, for $i = 1..n$, where $n \geq 0$, and result type $t$. Functions abstract over expressions.
9. For types $t_1, ..., t_n$, *connection[$t_1, ..., t_n$]* is the type of a connection (channel in π-calculus) which can send or receive values of types $t_1,...,t_n$ where $n \geq 0$.
10. For any types $t_1,...,t_n$, *abstraction[$t_1,...,t_n$]* is the type of an abstraction with parameter types $t_i$, for $i = 1..n$, where $n \geq 0$. Abstractions abstract over behaviours.

The world of data values is defined by the closure of rules 1 to 4 under the recursive application of rules 5 to 10.

## 6.2 Control constructs in the ArchWare ADL

The ArchWare ADL provides all of the usual control structures associated with expression-based languages, namely sequence, choice, iteration, and function call including recursion. To allow update change the ADL uses locations and assignment. Any data type may be stored in a location and be updated by a value of the same type.

Since the ArchWare ADL is formally based on the higher-order π-calculus it provides constructs analogous to those provided by the π-calculus for specifying control flow, communication and dynamic topology. The default execution pattern for behaviours in the ADL is parallel. In addition the ADL provides a rich set of control constructs.

Replication of a behaviour, indicated by ! in the π-calculus, is equivalent to a potentially infinite number of copies of that behaviour executing in parallel. This allows the specification of dynamic structure since replication generates copies as they are required. In Figure 4, the shown behaviour is replicated each time a value is received on connection *in_channel*. The behaviour waits at its reduction limit[1] for input. Upon receiving input it creates a clone of itself waiting at the reduction limit, and sends twice the received value on the *out_channel*. Many clones of the behaviour may be executing in parallel thus capturing dynamic topologies in the architecture, and supporting dynamic change.

```
replicate{
        via in_channel receive num ;
        via out_channel send 2 * num
     } ;
```

**Figure 4. Replication**

The *choose* clause, denoted by + in the π-calculus, allows the non-deterministic selection of one behaviour from two or more behaviours. In Figure 5, one of behaviours *client1*, *client2* or *client3* will be chosen at random by the run-time system.

---

[1] A behaviour reaches its reduction limit when it is waiting to communicate with another behaviour.

```
value client1 = … ;
value client2 = … ;
value client3 = … ;
choose  {    client1
     or      client2
     or      client3
} ;
```

**Figure 5. Choice**

Sequence, indicated by "." or *then* in the π-calculus, is denoted by ";" in the ADL. Therefore in Figure 4, *num* will be received on *in_channel* by the behaviour before the output value is sent on *out_channel*.

The π-calculus also provides the facility to restrict names to processes. In the ArchWare ADL this restriction is enabled partly by block structured programming scope rules and partly by an explicit *free* construct that specifies the values to be available for further binding.

### 6.3 Components

Software architectures describe systems in terms of their components and their interactions. Components are units of structure and functionality. In the ArchWare ADL components can be modelled by behaviours that are analogous to processes in the π-calculus. The code in Figure 4 specifies a server component that receives a number and sends back twice its value.

In order to facilitate design and reuse, the ADL allows the definition of abstractions that abstract over behaviours. Applying an abstraction results in a behaviour as illustrated in Figure 6.

```
value server = abstraction()
replicate
{    via in_channel receive num ;
     via out_channel send 2 * num } ;

server() ; ! applies the abstraction to yield a behaviour
```

**Figure 6. Abstraction**

There are two aspects to the interaction between components: coordination and communication. The former is concerned with synchronisation of components and the latter with exchange of data between components. Connections, analogous to channels in the π-calculus, are used for both aspects.

Behaviours can communicate, i.e. send and receive values, via connections if they share connections or if their connections have been explicitly unified. Empty messages via connections are used for coordination alone.

So far we have seen how to implement dynamic change through *replicate*, and update change through locations and assignment. We now turn our attention to the facilities for unexpected evolutionary change: hyper-code, decomposition, reflection and reification.

## 7 Hyper-code

The *hyper-code* abstraction was introduced in [25] as a means of unifying the concepts of source code, executable code and data in a programming system. The motivation is that this may ease the task of the programmer, who is presented with a simpler environment in which the conceptually unnecessary distinction between these forms is removed. In terms of Brooks' *essences* and *accidents*, this distinction is an accident resulting from inadequacies in existing programming tools; it is not essential to the construction and understanding of software systems [26]. In a hyper-code system the user composes hyper-code and the system executes it. When evolving the system, for example because an error has occurred, the user only ever sees a hyper-code representation of the program, which may now be partially executed. The hyper-code source representation of the program is structured and contains text and links to extant values.

Figure 7 shows an example of a hyper-code representation in the ArchWare ADL. The links embedded in it are represented by underlined tokens to allow them to be distinguished from the surrounding text. The first link is to an integer location value *count* that is used as a parameter in the application of the *server_abs* abstraction. The program also has two links to a previously defined abstraction *client_abs*. Hyper-code models sharing by permitting a number of links to the same value. Note that code values (*client_abs*) are denoted using exactly the same mechanism as data values (*count*). Note also that the value names used in this description have been associated with the values for clarity only, and are not part of the semantics of the hyper-code.

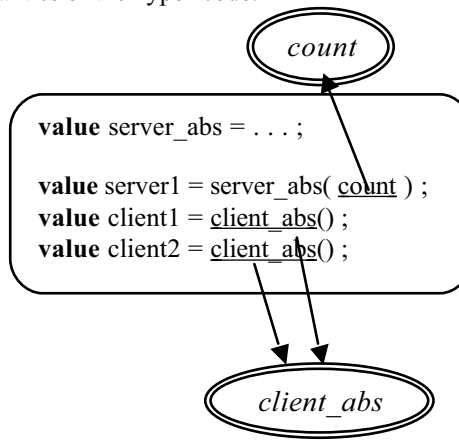

**Figure 7. ADL hyper-code**

The importance of hyper-code in active architectures is that it is rich enough to represent executing code. Thus as the program executes, the hyper-code changes in line with the semantics of the language. Since hyper-code can represent closure, through sharing links, it may be used as a representation for introspection of the executing system. Hyper-code provides the basic facility of preserving consistent state information across the system as it evolves.

There are a number of possible views of the concept of hyper-code. Most relevant to active architectures is the notion that hyper-code is an active executing graph with a programmable interface. In software engineering terms there is always a way of reverse engineering the executing computation so that it may be seen and programmed over.

## 8 Composition and decomposition

An essential property of evolutionary systems is the ability to decompose a running system into its constituent components, and compose evolved or new components to form a new system, while preserving any state or shared data if desirable.

The ArchWare ADL provides a *compose* operator which operates over a number of behaviours (components) and returns a single handle to these behaviours executing in parallel. The result of this composition is also a behaviour. Thus hierarchical systems may be modelled by components that are made up of other components. Figure 8 illustrates composition.

Behaviours *client* and *server* are composed to give *system*. The *as* construct permits meaningful labels to be associated with behaviours. As the details of *client* and *server* are not of interest in this example, the result value *position* sent by *server* is shown as a hyper-link which has previously been defined.

```
value channel_1 = connection() ;
value channel_2 = connection( string ) ;
value client = replicate{
               via channel_1 send ;
               via channel_2 receive pos : string } ;

value server = replicate{
                via channel_1 receive ;
                via channel_2 send position } ;

value system = compose{    pos_client as client
              and          pos_server as server }
```

**Figure 8. Composition**

The ADL also provides a *decompose* operator that breaks up a behaviour into its constituent behaviours. Decomposing a composite behaviour will undo any unification associated with the corresponding composition. On decomposition each constituent behaviour will eventually reach its reduction limit. Figure 9 illustrates the use of the decomposition operator on the composition from Figure 8.

Decomposition returns a sequence of views consisting of behaviours and their labels (if any) in the order that they were composed. All the behaviours are at their reduction limit for that composition. Figure 9 shows how these behaviours and their labels may be accessed from the sequence. These behaviours can be returned to the user as hyper-code, modified and recomposed.

```
value pos_seq = decompose system ;

value client_val = pos_seq::1.bhvr ;
value server_val = pos_seq::2.bhvr ;
value comp1_label = pos_seq::1.label
```

**Figure 9. Decomposition**

Explicit composition is not always required since two behaviours will communicate if they share the same connection and communication is ready. Composition gives a handle to the new composed behaviours. The higher order nature of the language means that two behaviours that have been sent to a third may wish to communicate but do not share the same connection value. The *compose* operation has a variant that allows the unification of connection values during composition to facilitate communication in these circumstances. Figure 10 shows an example of unification.

```
value client =    abstraction()
{  value out_request = connection() ;
   value in_reply = connection( string ) ;
   replicate{
            via out_request send ;
            via in_reply receive pos : string }
} ;

value server = abstraction()
{  value in_request = connection() ;
   value out_reply = connection(string) ;
   replicate{
            via in_request receive ;
            via out_reply send position }
} ;

value system =
   compose{  pos_client as client() and
             pos_server as server()
      where {  pos_client::out_request unifies
                  pos_server::in_request,
               pos_client::in_reply unifies
                  pos_server::out_reply } }
```

**Figure 10. Unification**

## 9 Reflection and reification

A hyper-code system may be thought of as operating within two abstract domains: entities (E) and representations (R). E contains all the first class values defined by the language while R contains the concrete representations of the values in E. Given these domains, four domain operations over E and R may be defined.

- *reflect* maps a representation to its corresponding entity (R⇒E)
- *reify* maps an entity to a corresponding representation (E⇒R)
- *execute* executes an entity, possibly generating a result (E⇒E)
- *transform* maps one representation to another (R⇒R)

Reflection and reification are of particular interest since they play a vital role in evolving systems. Once a system is decomposed, reification allows us to view representations of its components. These may be evolved to capture new requirements. Reflection allows evolved or new components to be bound back into the system.

The ability of hyper-code to capture closures allows us to represent parts of a system after decomposition without losing their context. It provides representations that can be used for both evolving the components and recomposing them into the new system.

The ArchWare ADL provides specific operations for *reflect* and *reify*. Evolution is effected by decomposing the selected part of the system, reifying the components, applying a transformation and finally reflecting the code. The specific evolutionary process may be abstracted by writing a style to specify the evolution and the constraints upon it.

## 10 An example of evolution

The concepts discussed in earlier sections are now illustrated using a (by necessity simple) example written in the ArchWare ADL. This example is based on a long running *in silico* experiment [27]. The user is a scientist who wants to run the experiment from a desktop-based client. The experiment itself runs on a server machine with access to a range of resources, e.g. corporate databases. The scientist's client will connect to a server that disseminates data about the status of the experiment. The client also allows the scientist to control (start and end) the experiment. Initially there will be a single client and a single server, but later the system will be evolved so that the functionality of the server is split into two.

The functionality of the server and the client can be modelled as abstractions in the ArchWare ADL. When applied, these abstractions yield executing behaviours. Such behaviours are the components that make up the client-server system. The repetitive nature of both the client and the server is captured using replication. Thus the dynamic nature of the system is already present in that the server may replicate itself to deal with data from the experiment and the client may replicate itself in parallel to react to the data sent by the server.

Since exact details of the experiment are not of interest to this example, we will assume that values of data type *exp_view* provide all necessary information about the experiment. The client abstraction can then be defined as shown in Figure 11.

```
! client
value client_abs = abstraction()
{   value c_start = connection() ;
    value c_stop = connection() ;
    value c_get = connection( exp_view ) ;
    via c_start send ;
    replicate
        choose{
            {   via c_get receive ev : exp_view ;
                via c_display send ev } or
            {   via user_input receive ;
                via c_stop send } }
}
```

**Figure 11. The client abstraction**

The client defines the connections it needs to communicate on, sends a message to start the experiment and then on demand replicates itself to choose either to receive details of the experiment and display it to the user or to receive a command from the user and send a message to end the experiment. The former provides the scientist with an ongoing view of the experiment and the latter with the means to stop the experiment if its progress is not satisfactory. *c_display* and *user_input* connections are shown as hyper-links in the code as they have previously been defined elsewhere.

Figure 12 shows the definition of the server abstraction. The body of the server mirrors that of the client. It defines its connections, receives the start message, begins the experiment and then on demand replicates itself to choose either to receive the stop message and end the experiment or to receive the current values of the experiment and send them on. As before connection *exp_input* and function *stop_experiment* are shown as hyper-links.

```
! server
value server_abs = abstraction()
{   value s_start = connection() ;
    value s_stop = connection() ;
    value s_put = connection( exp_view ) ;
    via s_start receive ;
    start_experiment() ;
    replicate
        choose{
            {   via s_stop receive ;
                stop_experiment() } or
            {   via exp_input receive current_view ;
                via s_put send current_view } }
}
```

**Figure 12. The server abstraction**

Having defined server and client abstractions, we can now create a client-server system by composing instances of the server and the client abstractions with appropriate unification. Unification ensures that corresponding client and server connections are matched for communication. Defining the composition as a value gives us a handle (*CS_system1*) to the resulting behaviour. Figure 13 shows the composition of one client and one server.

The *as* construct allows users to associate meaningful labels with behaviours being composed. In addition to aiding the identification of behaviours after decomposition, this facility also connections to be uniquely identified for unification.

```
! client-server system
value CS_system1 =
compose{
        client as client_abs() and server as server_abs()
        where{   client::c_start unifies server::s_start,
                 client::c_stop unifies server::s_stop,
                 client::c_get unifies server::s_put }
     }
```

**Figure 13. The client-server system**

Once the system starts executing, we may wish to change its structure. The scientist may want to share a view of the *in silico* experiment with colleagues, or the experiment may take longer than expected and the scientist may wish to get advice before deciding whether the server should be stopped, aborting the experiment.

We begin this process by decomposing the system into its component parts as shown in Figure 14. The result of this decomposition is a sequence of views containing the following information about each behaviour of the system: label, behaviour value and list of connections. In long-running systems, labels associated with behaviours may help identify their purpose and identity.

```
! decompose system
value cs_seq = decompose CS_system1
```

**Figure 14. Decomposition**

Necessary changes can then be made by evolving or redefining some components. In this case we wish to split the functionality of the server into two by creating two new servers, one serving status alone, which can be shared among multiple clients, and the other serving the command messages, of which the scientist who started the experiment wants to retain control. Therefore we create two new abstractions to replace the old *server_abs*.

Using hyper-code representations of the abstractions will enable us to define the new abstractions to use the current values of variables without having to explicitly store and reinitialise them as shown in Figure 15. Abstraction *view_server_abs* disseminates status information about the experiment while *command_server_abs* allows the experiment to be controlled. Note that start messages are ignored as the experiment has already been running.

```
! view server
value view_server_abs = abstraction()
replicate
{   via exp_input receive current_view ;
    via s_put send current_view
}

! command server
 value command_server_abs = abstraction()
replicate
   choose{
       {    via s_start receive }
      or
       {   via s_stop receive ;
          stop_experiment() }
```

**Figure 15. The new server abstractions**

A new client-server system can then be formed by composing the two new servers with the decomposed client appropriately as shown in Figure 16.

```
! make new client-server system
value CS_system2 =
compose{ client as cs_seq::1.bhvr
    and    view_server as view_server_abs()
    and    command_server as command_server_abs()
    where{
        client::c_start unifies command_server::s_start,
        client::c_stop unifies command_server::s_stop,
        client::c_get unifies view_server::s_put }
} ;
```

**Figure 16. The evolved system**

Now the client will communicate with both servers. If experiment information is required the client will talk to *view_server* and if the scientist wishes to control the experiment then the client will talk to *command_server*.

## 11 Conclusions

This paper describes the ArchWare Architecture Description Language, which is sufficiently rich to provide executable specifications of active systems. We define a core language on which architectural styles can be layered and on which a construction methodology can be applied. We focussed on the base technologies required to support evolvable systems and presented examples of how the ADL may be used to model active architectures. In particular, we showed how architectures specified in the ADL were dynamic, updateable, decomposable and reflective.

We postulate that the need for dynamic evolution of software architecture definitions is inherent in the description of autonomic systems and will give a paradigm for architectural transformations. Furthermore we recognise the role that hyper-code may have in reverse engineering. For reliability it is important that the architectural definition of a system is automatically kept consistent with the state of the system at all times during its execution and evolution. Our approach to meeting these requirements involves the combination of a number of technologies: reflection, reification and representation for closure (hyper-code).

We have combined these with a variant of the π-calculus that yields dynamic expressions and communication while providing the basis for formal analysis tools in the form of theorem provers, type checkers and model checkers.

These elements provide a core evolutionary support system. At the time of writing the core ArchWare ADL, the hyper-code, the decomposition operator and the reflective mechanism are all implemented and being used to specify evolvable software architectures.